\documentclass[11pt,british]{paper}
\usepackage[T1]{fontenc}
\usepackage[latin9]{inputenc}
\usepackage{geometry}
\usepackage{amsmath}
\usepackage{esint}
\usepackage[numbers]{natbib}

\usepackage{color}
\definecolor{ngreen}{rgb}{0.2,0.6,0.2}
\definecolor{red}{rgb}{0.8,0.2,0.2}

\makeatletter
\newcommand{\lyxaddress}[1]{
\par {\raggedright #1
\vspace{1.4em}
\noindent\par}
}

\makeatother

\usepackage{babel}
\begin{document}

\title{Bell nonlocality, signal locality and unpredictability (or What Bohr
could have told Einstein at Solvay had he known about Bell experiments)}

\author{Eric G. Cavalcanti $^{1,2}$ and  Howard M. Wiseman $^{2}$}
\maketitle

\lyxaddress{$^{1}$School of Physics, The University of Sydney, Sydney NSW 2006, Australia\\
$^{2}$Centre for Quantum Dynamics, Griffith University, Brisbane QLD 4111, Australia}

\begin{abstract}
The 1964 theorem of John Bell shows that no model that reproduces
the predictions of quantum mechanics can simultaneously satisfy the
assumptions of locality and determinism. On the other hand, the assumptions
of \emph{signal locality} plus \emph{predictability} are also sufficient
to derive Bell inequalities. This simple theorem, previously noted
but published only relatively recently by Masanes, Acin and Gisin, has fundamental
implications not entirely appreciated. Firstly, nothing can be concluded
about the ontological assumptions of locality or determinism independently
of each other -- it is possible to reproduce quantum mechanics with
deterministic models that violate locality as well as indeterministic
models that satisfy locality. On the other hand, the operational assumption
of signal locality is an empirically testable (and well-tested) consequence
of relativity. Thus Bell inequality violations imply that we can trust
that some events are fundamentally \emph{unpredictable}, even if we
cannot trust that they are indeterministic. This result grounds the
quantum-mechanical prohibition of arbitrarily accurate predictions
on the assumption of no superluminal signalling, regardless of any postulates of
quantum mechanics. It also sheds a new light on an early stage of
the historical debate between Einstein and Bohr.
\end{abstract}

\section{Introduction}

Bell's seminal 1964 paper~\citep{Bell1964} shows that quantum correlations
violate the conjunction of locality\footnote{Our usage of `locality' here is the same as that of Bell in 1964: ``that
the result of a measurement on one system be unaffected by operations
on a distant system''. This assumption is sometimes referred to as
``parameter independence'' and is strictly weaker than the assumption
of ``local causality'', which was later introduced by Bell~\citep{Bell1976} 
to show that determinism does not need to be assumed, and that no
locally causal model (deterministic or otherwise) can reproduce quantum
mechanics. See also Sec.~\ref{sec:Conclusion}.} and determinism. However, there are quantum models that violate locality
but maintain determinism (Bohmian mechanics \citep{Bohm1952I} is
an example), and models that maintain locality but violate determinism
(standard operational quantum theory is an example). Thus nothing
can be concluded from Bell's theorem about locality or determinism
independently of each other.

Here we show that a remarkable conclusion can be reached by deriving
Bell inequalities from a different set of assumptions: \emph{signal
locality }(i.e. the impossibility to send signals faster than light)
and \emph{predictability} (i.e. the assumption that one can predict
the outcomes of all possible measurements to be performed on a system).
These assumptions are purely \emph{operational}, that is, they refer
to operational quantities only, in contrast to the \emph{ontological}
assumptions of locality and determinism, which refer to properties
of a hidden-variable model that reproduces the observations. In particular,
signal locality is an uncontroversial consequence of relativity, as
opposed to locality, which can be violated by the underlying hidden-variable
model while still maintaining signal locality at the observable level
(as is the case with Bohmian mechanics, which reproduces quantum mechanics). 

This derivation therefore allows us to confidently conclude that \emph{predictability}
must fail for experiments that violate Bell inequalities. In other
words, it allows the conclusion that, if it is impossible to signal
faster than light, then it is impossible to predict the outcomes of
experiments that violate Bell inequalities, even if those outcomes
might be determined by an underlying hidden-variable model.

\medskip{}

This work builds on and clarifies some recent results. In the 1994
paper where they suggested Bell nonlocality as a ``natural'' axiom for quantum theory and 
brought to light the existence of stronger-than-quantum
correlations compatible with signal locality, Popescu and Rohrlich
\citep{Popescu1994b} commented on an unpublished result of Aharonov
to the effect that ``relativistic causality'' and ``nonlocality''
imply ``indeterminism''. Here we would like to point out that, strictly speaking, this claim is incorrect:
the conjunction of signal locality (i.e., what they meant by ``relativistic
causality'' and Bell-inequality violation (what they meant by ``nonlocality'')
implies the failure of predictability, but not of determinism. While
Aharonov's result was presumably correct with the appropriate translation,
the choice of words seems to imply they did not make at that stage
the distinction between determinism and predictability, and this work
should clarify the importance of doing so.

This result also relates to the mechanism underlying the security
of quantum key distribution based on signal locality and Bell inequality
violations, as proven by Barrett, Hardy and Kent \citep{Barrett2005a}.
In the discussion of their result, they cited a paper of Valentini
\citep{Valentini2002} that purported to show that ``any state that
is deterministic and nonlocal allows signalling''. Since what they
meant by ``determinism'' was the same as we here mean by ``predictability'',
this is essentially the contrapositive of the present result. However,
this was strictly speaking not proven in Valentini's paper. What he
claimed to have proven was that for all nonlocal deterministic hidden-variable
theories, a violation of signal locality occurs if and only if the
theory allows a distribution of hidden variables different from that
which is needed to reproduce quantum mechanics, i.e., if and only
if the distribution is different from that of ``quantum equilibrium''%
\footnote{Besides, Valentini's main result seems to be flawed. He seems to have
only proven the weaker result that \emph{there exist} distributions
over the hidden variables which would allow signalling, as the reader
may be convinced by analysing the first equation on page 276 of \citep{Valentini2002}.%
}. 

To our knowledge, the first correct published proof of this result
can be found in an article by Masanes, Acin and Gisin, that, following the suggestion of Popescu and Rohrlich, studied general properties of non-signalling theories \citep{Masanes2006General}. However, they also
did not distinguish between predictability and
determinism (like Aharonov and Valentini, they used ``determinism'' to
refer to what we call ``predictability''). The present discussion should
thus serve to clarify the conceptual basis and importance of this
result%
\footnote{After a first version of this work was posted on the arXiv (arXiv:0911.2504v1),
another arXiv post (arXiv:0911.3427v1, eventually published in Nature
\citep{Pironio2010Random}) underscored the importance of the distinction
by proposing a scheme to generate random numbers certified by violation
of a Bell inequality. Those results are however somewhat distinct
from the present one, in that to derive bounds on the randomness of
the output of a Bell experiment, those authors assumed the validity
of the laws of quantum mechanics. Here no such assumption is made
(but consequently no bound is given on the randomness or unpredictability
of the output).%
}.

\medskip{}

This result also has an interesting didactical implication for the
famous dialogues between Einstein and Bohr on the foundations of quantum
theory at the 1927 and 1930 Solvay conferences. At this stage, prior to the 1935 
Einstein, Podolsky
and Rosen paper \citep{Einstein1935} and the concept of entanglement
that was born from it, it seems that 
Einstein attempted to attack the \emph{validity
}of quantum mechanics, not only its \emph{completeness}, by concocting
thought experiments aimed at obtaining a violation of the uncertainty
principle~\citep{BohrEinst}. However, by carefully applying the
uncertainty principle to the experimental apparatuses as well as the
systems being measured, Bohr showed that the uncertainty principle
was \emph{consistent}; that is, he showed that \emph{if }the uncertainty
principle is valid for the degrees of freedom of all measuring apparatuses
\emph{then} those measuring apparatuses can't be used to violate the
uncertainty principle associated to a quantum system. Interestingly,
in the last of such attempts from Einstein, Bohr used Einstein's own
theory of general relativity to demonstrate the consistency of quantum
mechanics. Could Bohr have gone beyond that, and argued, with appeal
to independent fundamental principles, that the uncertainty principle
\emph{must} be valid? 

In the remainder of this paper we will answer this question in the
affirmative, and show that an uncontroversial consequence of Einstein's
theory of special relativity (signal locality) and some raw experimental
observations (namely, the violation of Bell inequalities) lead to
a weak version of the uncertainty principle---perfect predictability
of natural phenomena must be impossible, regardless of any of the
postulates of quantum mechanics.

This paper is structured as follows. In Section \ref{sec:Experimental-metaphysics}
I introduce the basic concepts and notation required for the main
result. The main result is proven in Section \ref{sec:Bell,signal_locality},
followed by a discussion of the result and concluding remarks in Section
\ref{sec:Conclusion}.

\section{Experimental metaphysics\label{sec:Experimental-metaphysics}}

Abner Shimony coined the term ``experimental metaphysics'' to refer
to the field of study pioneered by Bell, where general metaphysical%
\footnote{For the physicist trained to be suspicious of philosophical terms,
note that in this context the term `metaphysics' does not refer to
mysticism, but to the study of formal and empirical properties of
physical theories themselves. (Experimental) metaphysics is to physics
as metamathematics is to mathematics. It includes the study of sets
of physical theories which \emph{fail} to represent observations,
where this analysis can be illuminating in understanding those that
\emph{do not}.%
} concepts such as ``local causality'' are shown to lead to experimental
constraints which can be tested in the laboratory. In the following
we will introduce the concepts required to prove our main result.

The experimental setup considered here involves two spatially separated
observers, Alice and Bob, who can perform a number of measurements
and observe their outcomes. For each pair of systems they perform
measurements upon, the choices of measurement settings and their respective
outcomes occur in regions which are space-like separated from each
other, so that no signal travelling at a speed less than or equal
to that of light could connect any two of them. For each pair of systems,
we will denote by $a$ and $b$ Alice's and Bob's respective measurement
settings, and by $A$ and $B$ their corresponding observed outcomes. 
Note that here we are following the notational convention Bell established in 1964 \cite{Bell1964}. 
Each pair of systems is prepared by an agreed-upon reproducible procedure
$\kappa$ (which in quantum mechanics would define a quantum state
for the pair of systems). 

We will define a \textbf{phenomenon},\emph{ }for a given preparation
procedure $\kappa$, by the relative frequencies
\begin{equation}
f(A,B|a,b,\kappa).\label{eq:phenomenon}
\end{equation}
for all measurements $a$, $b$, and corresponding outcomes $A$,
$B$. Note that this definition does not assume a frequentist interpretation
of probabilities. It simply acknowledges that in any physical experiment,
the actual phenomenon observed is encoded by those relative frequencies
(with some associated statistical uncertainty that can in principle
be made arbitrarily small). When an equation involving variables appears,
it is to be understood that the equality holds for all values of those
variables.

We will say that a phenomenon is \textbf{predictable}, or that it
satisfies \textbf{predictability} if and only if 
\begin{equation}
f(A,B|a,b,\kappa)\in\{0,1\}.\label{eq:predictability-1}
\end{equation}
Now of course it is a consequence of the postulates of quantum mechanics
that there are unpredictable phenomena, according to Heisenberg's uncertainty principle. However, we will see that this conclusion can
be reached independently of the postulates of quantum theory, by using
\textbf{signal locality}, i.e. the assumption that 
\begin{equation}
f(A|a,b,\kappa)=f(A|a,\kappa),\label{eq:SL-1}
\end{equation}
and the corresponding equation for $B$. Note that all the definitions
above are purely operational, i.e. they refer to operationally defined,
observable quantities only.

What Bell did was to consider, as did Einstein, Podolsky and Rosen,
the possibility that there might be further variables (in addition
to $a$, $b$ and $\kappa$) that are relevant to the phenomenon observed.
We represent any such variables by the symbol\textbf{ }$\lambda$%
\footnote{Note that considering the possibility that further variables exist
is not the same as assuming that they exist; there is no ``hidden-variable
assumption'' in Bell's theorem.%
}. They are not fully determined by the preparation procedure $\kappa$,
and as such may be deemed ``hidden variables''. A more appropriate
terminology is ``ontic variables'', since they represent any ``real''
physical state of the parts of the world which are relevant to the
experiments being considered. This terminology also emphasises the
distinction between $\kappa$ and $\lambda$: $\kappa$ represents
the relevant variables \emph{known }by Alice and Bob (or by whichever
party is describing the phenomenon), and $\lambda$ represents the
variables that are objectively relevant to the experiments considered,
regardless of whether they are known or even knowable%
\footnote{Indeed it is a corollary of the present result that in any deterministic
model that reproduces quantum theory the ontic variables must be necessarily
unknowable.%
}.

An \textbf{ontological model} \citep{Spekkens2005,Rudolph2006,Harrigan2007b,Harrigan2007c}
for a phenomenon is one in which the phenomenon can be explained by
considering ontic variables.\emph{ }It consists of the set $\Lambda$
of values of $\lambda$, together with a probability density $\mu(\lambda|\kappa)$
for every preparation procedure $\kappa$ and a specification of \emph{
\begin{equation}
P(A,B|a,b,\kappa,\lambda)\label{eq:modelprob}
\end{equation}
}which reproduces the phenomenon by
\begin{equation}
\int_{\Lambda}d\lambda\,\mu(\lambda|\kappa)\, P(A,B|a,b,\kappa,\lambda)=f(A,B|a,b,\kappa).\label{eq:modelfreq}
\end{equation}

Note that in Eq. \eqref{eq:modelfreq} we have used the assumption
of \textbf{free variables}, which\emph{ }is the assumption that the
choices of experiment $a$, $b$, can be conditioned on variables
which are uncorrelated with $\lambda$. Formally, this is the assumption
that\emph{ 
\begin{equation}
\mu(\lambda|a,b,\kappa)=\mu(\lambda|\kappa).\label{eq:freewill}
\end{equation}
}This is sometimes called the \emph{``}free will''\emph{ }assumption,
but that terminology seems to imply something about human capabilities
that doesn't seem to be necessary for the purposes at hand.

We are now ready to define properties of models. A model is said to
satisfy \textbf{locality} if and only if 
\begin{equation}
P(A|a,b,\kappa,\lambda)=P(A|a,\kappa,\lambda),\label{eq:L}
\end{equation}
plus the corresponding equation for $B$\footnote{We remind the reader that this corresponds 
to Bell's definition of locality introduced in 1964 \cite{Bell1964}. There are instances in less formal publications, for example Ref.~\cite{Bell-Proc76}, where Bell used the term ``locality'' more loosely to mean the property which quantum mechanics lacked, as 
revealed by his theorem. However these are very much the exception, 
and from 1976 on, Bell almost invariably used the term ``local causality'' for this property.}.    A model is said to satisfy
\textbf{signal locality} if and only if 
\begin{equation}
P(A|a,b,\kappa)=P(A|a,\kappa),\label{eq:SL}
\end{equation}
plus the corresponding equation for $B$. Note that the left-hand-side of the equation here 
is defined within the model as the left-hand-side of Eq.~(\ref{eq:modelfreq}). 
We say that locality is
an \emph{ontological }concept because it refers to ontic variables
in its definition\emph{, }while signal locality is an \emph{operational
}concept\emph{ }because it only refers to operational variables in
its definition. Note that a model satisfies signal locality if and
only if the corresponding phenomenon also does, i.e., if and only
if $f(A|a,b,\kappa)=f(A|a,\kappa)$. 

To see that a violation of Eq. \eqref{eq:SL} would imply the possibility
to transmit signals between the experimental sites, note that if the
phenomenon violates signal locality, then there exist at least two
possible choices of setting $b$, $b'$ such that $f(A|a,b,\kappa)\neq f(A|a,b',\kappa)$.
Therefore by looking at the frequency of outcomes of $A$ in a large
enough ensemble (and in principle it is possible for Alice to make
all of the measurements in her ensemble space-like separated from
all measurements in Bob's ensemble), Alice can determine with arbitrary
accuracy what setting Bob has chosen, thus allowing Bob to send signals
to Alice. 

Violation of \emph{locality}, on the other hand, does not imply signalling,
since in general only the probabilities for Alice's outcomes \emph{conditioned
on the hidden variables} depend on the choice of experiments at Bob's
site. But since those hidden variables can be unknowable \emph{in
principle}, that kind of non-locality cannot necessarily be used to
transmit signals. Bohmian mechanics is an example of a model that
violates locality but not signal locality.

We now come to the crucial distinction between the concepts of \emph{determinism}
and \emph{predictability}. A model is said to be \textbf{deterministic},
or to satisfy \textbf{determinism}%
\footnote{There is another useful sense of determinism which needs to be distinguished
from the one we  are using here. Quantum mechanics can be said to be
deterministic in the sense that for a closed system the state at a
later time is determined through unitary evolution by the state at
an initial time. However, operational quantum mechanics is not deterministic
in the sense used in this paper, since of course a system undergoing
a measurement interaction is no longer a closed system.%
}, if and only if 
\begin{equation}
P(A,B|a,b,\kappa,\lambda)\in\{0,1\},\label{eq:determinism}
\end{equation}
which implies that $A$ and $B$ can be specified as functions as follows: 
\begin{equation}
A=A(a,b,\kappa,\lambda),\; B=B(a,b,\kappa,\lambda).\label{eq:deterministicfunctions}
\end{equation}
On the other hand, a model is said to be \textbf{predictable}, or
to satisfy \textbf{predictability} if and only if 
\begin{equation}
P(A,B|a,b,\kappa)\in\{0,1\}.\label{eq:predictability}
\end{equation}
This implies that $A$ and $B$ can be specified as functions as follows: 
\begin{equation}
A=A(a,b,\kappa),\; B=B(a,b,\kappa).\label{eq:predictablefunctions}
\end{equation}
Determinism, like locality, is an ontological concept while predictability, like signal locality, is an operational concept. As with signal locality, a model is predictable if and only
if the phenomenon it reproduces is predictable. It is interesting
to note that it is impossible for a phenomenon to violate determinism
by itself: every phenomenon can be given a deterministic model, simply
by postulating a sufficient number of hidden variables.

Obviously, predictability implies determinism, but the converse is
not true. Heisenberg's uncertainty principle implies that no model
that reproduces quantum mechanics is predictable, but it imposes no
limitation on determinism. In fact, there are models of quantum mechanics
(Bohmian mechanics is an example) which are deterministic, but the
following theorem shows that they must nevertheless be unpredictable
(or violate signal locality).

\section{Bell inequalities from signal locality and predictability\label{sec:Bell,signal_locality}}

John Bell's 1964 theorem \citep{Bell1964} demonstrated that the conjunction
of the concepts of \emph{locality }and \emph{determinism }as defined
above leads to a set of experimental constraints known as \emph{Bell
inequalities}, and that some predictions of quantum mechanics regarding
entangled states violate those inequalities. For present purposes
we just need to point out that to derive a Bell inequality it is sufficient
to require that the joint probabilities of experimental outcomes given
by a model is \textbf{factorisable}, i.e., that it can be written
in the form
\begin{equation}
P(A,B|a,b,\kappa,\lambda)=P(A|a,\kappa,\lambda)P(B|b,\kappa,\lambda).\label{eq:factorisability}
\end{equation}
This equation is one way of expressing the condition on which Bell bestowed the name  
{\em local causality} \cite{Bell1976}. 

To see that locality and determinism imply factorisability, note that
the joint probabilities of any model can be written as $P(A,B|a,b,\kappa,\lambda)=P(A|B,a,b,\kappa,\lambda)P(B|a,b,\kappa,\lambda).$
If the model is deterministic, then $P(A|B,a,b,\kappa,\lambda)=P(A|a,b,\kappa,\lambda)$,
because $A$ is already determined by $a,b,\kappa,\lambda$. From
the definition of locality given by Eq.~\eqref{eq:L} we thus arrive
at the factorisable model of Eq.~\eqref{eq:factorisability}.

We now arrive at our main result, that the conjunction of signal locality
and predictability also allow one to derive Bell inequalities. The
proof is simple. The definition of predictability (Eq.~\eqref{eq:predictability}),
implies that 
\begin{equation}
P(A,B|a,b,\kappa,\lambda)=P(A,B|a,b,\kappa)=P(A|B,a,b,\kappa)P(B|a,b,\kappa).\label{eq:pred}
\end{equation}
The first equality follows because the second expression must be
either $0$ or $1$, thus those probabilities cannot be altered by conditioning
on $\lambda$. The second equality follows from the definition of conditional
probabilities. Eq.~\eqref{eq:predictability} also implies that $P(A|B,a,b,\kappa)=P(A|a,b,\kappa)$,
since $A$ is already specified by $a,b,\kappa$. Thus $P(A,B|a,b,\kappa,\lambda)=P(A|a,b,\kappa)P(B|a,b,\kappa).$
Assuming signal locality, i.e.~Eq.~\eqref{eq:SL}, we obtain 
\begin{equation}
P(A,B|a,b,\kappa,\lambda)=P(A|a,\kappa)P(B|b,\kappa).\label{eq:main_result}
\end{equation}
This has the factorisable form of \eqref{eq:factorisability}, as
we set out to prove. To make that more explicit, note that this is
equivalent to a model with $P(A|a,\kappa,\lambda)=P(A|a,\kappa)$
and $P(B|b,\kappa,\lambda)=P(B|b,\kappa)$ for all $\lambda.$ Therefore,
signal locality plus predictability imply factorisability, and thus
Bell inequalities.

\section{Discussion and conclusion\label{sec:Conclusion}}

Bell was adamant in stressing that his concept of locality was distinct
from the concept of signal locality.  
In fact, he rejected the importance
of the concept of signal locality, as he understood that it was hard
to talk about it without using apparently anthropocentric terms like
`information' and `controllability':
\begin{quote}
\textquotedbl{}Suppose we are finally obliged to accept the existence
of these correlations at long range {[}...{]}. Can \emph{we} then
signal faster than light? To answer this we need at least a schematic
theory of what \emph{we} can do, a fragment of a theory of human beings.
Suppose we can control variables like $a$ and $b$ above, but not
those like $A$ and $B$. I do not quite know what `like' means here,
but suppose the beables somehow fall into two classes, `controllables'
and `uncontrollables'. The latter are no use for \emph{sending} signals,
but can be used for \emph{reception.}\textquotedbl{} \citep{Bell1976}.
\end{quote}
And he rejects the idea that signal locality could be taken as the
fundamental limitation imposed by relativity:
\begin{quote}
\textquotedbl{}Do we have to fall back on `no signalling faster than
light' as the expression of the fundamental causal structure of contemporary
theoretical physics? That is hard for me to accept. For one thing
we have lost the idea that correlations can be explained, or at least
this idea awaits reformulation. More importantly, the `no signalling...'
notion rests on concepts which are desperately vague, or vaguely applicable.
The assertion that `we cannot signal faster than light' immediately
provokes the question:
\begin{quote}
Who do we think \emph{we} are?
\end{quote}
\emph{We} who can make `measurements', \emph{we} who can manipulate
`external fields', \emph{we }who can signal at all, even if not faster
than light? Do \emph{we} include chemists, or only physicists, plants,
or only animals, pocket calculators, or only mainframe computers?\textquotedbl{}
\cite{Bell1990}.
\end{quote}
Bell is right in that we cannot define signal locality without referring
to operational or epistemic concepts, but we find this situation more acceptable  
than Bell did. The reason is that `information' and thus `signal'
\emph{are} themselves operational concepts. We do indeed need to know
which variables are controllable, and which variables are knowable,
in order to define signal locality. However, while ``controllable''
and ``knowable'' may be as philosophically problematic as ``information'',
they are also just as pragmatically clear and useful. Besides, the
use of these terms does not need to imply an anthropocentric view
of physics as he seemed to be worried in the passage above. There
seems to be no fundamental difficulty in talking about, say, \emph{machines}
``knowing'' or ``controlling'' the variables defined in the previous
section.

\medskip{}

Since experiments routinely violate Bell inequalities (up to some
open loopholes), we can conclude, through the theorem proven in the
previous section, that signal locality and predictability cannot both
the true. Now this allows an interesting conclusion. Bell showed in
1964 that locality and determinism cannot both be true. Some people
have taken that to imply indeterminism, choosing to keep locality.
However, a deterministic model (Bohmian mechanics) exists that reproduces
quantum theory, while violating locality. One may reject this or similar
models on other grounds (e.g.~elegance, symmetry, etc.), but Bell's
theorem cannot be used to that end. Furthermore, Bell later showed
that the stronger concept of local causality also implies Bell inequalities
and must therefore be false. So even if one chooses to reject determinism,
the resulting indeterminism still has a nonlocal character, by Bell's
later argument. The result is thus that nothing can be concluded separately
about the ontological claims of determinism and locality from the
violation of Bell inequalities.

However, the validity of signal locality has a much less controversial
footing. Even those who believe in the violation of locality would
generally agree that the operational assumption of signal locality
must be valid, and that while a violation of locality may have a ``peaceful
coexistence'' \cite{Shi89} with relativity, a violation of signal locality would
be in direct contradiction with it. Therefore the full weight of the
violation of Bell inequalities can be confidently transferred to the
violation of the operational concept of predictability: there can
be no predictable model that allows violation of a Bell inequality.
To the extent that those violations actually occur in nature, we can
conclude that the world is indeed fundamentally unpredictable. 

Furthermore, we arrive at this conclusion without needing to assume
anything about quantum mechanics. It is simply a consequence of bare
experimental data and an uncontroversial consequence of the theory
of relativity. Bohr, of course, could not have known about this result
before Bell, but if he did, he could have much more easily convinced
Einstein that his attempts to make quantum phenomena predictable were
bound to fail%
\footnote{Although, of course, this would have no implication for Einstein's
later attacks on the \emph{completeness} of quantum theory. Of course all that we have said about the Bohr-Einstein debates accepts 
 that Bohr's narrative of the events is accurate. It has been argued by Howard \cite{How89} and by Bacciagaluppi and Valentini 
\cite{BV09} that Bohr misunderstood Einstein's arguments, and that actually Einstein was arguing for incompleteness (based on 
nonseparability), rather than incorrectness, the whole time.}. And in another twist of irony, he could have again used Einstein's
own theory of relativity to prove him wrong.

\section*{Acknowledgments} 

We are %also 
grateful to Rob Spekkens
for useful feedback and encouragement. This work was partly supported
by the Australian Research Council Discovery grant DP0984863 and Discovery Early-Career
Researcher Award DE120100559.

{\footnotesize \bibliographystyle{plain}

\begin{thebibliography}{10}

\bibitem{BV09}
G.~Bacciagaluppi and A.~Valentini.
\newblock {\em Quantum theory at the crossroads : reconsidering the 1927 Solvay conference}.
\newblock Cambridge University Press, 2009.

\bibitem{Barrett2005a}
J.~Barrett, L.~Hardy, and A.~Kent.
\newblock {No signaling and quantum key distribution}.
\newblock {\em Physical Review Letters}, 95(1):010503, 2005.

\bibitem{Bell1964}
J.~S.~Bell.
\newblock On the {E}instein-{P}odolsky-{R}osen paradox.
\newblock {\em Physics}, 1:195, 1964.

\bibitem{Bell-Proc76}
J.~ S.~Bell.
\newblock {Einstein-Podolsky-Rosen Experiments}.
\newblock {\em Proceedings of the symposium on Frontier Problems in High Energy Physics,} 33-45. 
\newblock Pisa, June 1976. Reproduced in Ref.~\cite{Bell2004}.

\bibitem{Bell1976} 
J.~S.~Bell. 
\newblock The Theory of Local Beables 
\newblock {\em Epistemological Lett.} 9, 1976. Reproduced in Ref.~\cite{Bell2004}.

\bibitem{Bell1990} 
J.~S.~Bell. 
\newblock La nouvelle cuisine 
\newblock in {\em Between Science and Technology}, eds. A. Sarlemijn and P. Kroes.  Elsevier Science Publishers, 1990. Reproduced in Ref.~\cite{Bell2004}.

\bibitem{Bell2004}
J.~S.~Bell.
\newblock {\em {Speakable and Unspeakable in Quantum Mechanics}}, 2nd Ed.
\newblock Cambridge University Press, 2004.

\bibitem{Bohm1952I}
David Bohm.
\newblock {A Suggested Interpretation of the Quantum Theory in Terms of
  "Hidden" Variables. I}.
\newblock {\em Physical Review}, 85(2):166--179, 1952.

\bibitem{BohrEinst} 
N.~Bohr.
\newblock {Discussion with Einstein on epistemological problems in atomic physics}.
in Ref.~\cite{Sch49}, pp. 201--241; reproduced in \cite{QTMeas1983}.

\bibitem{Einstein1935}
A.~Einstein, B.~Podolsky, and N.~Rosen.
\newblock {Can quantum-mechanical description of physical reality be considered
  complete?}
\newblock {\em Physical Review}, 47:777, 1935.

\bibitem{Harrigan2007c}
N.~Harrigan and T.~Rudolph.
\newblock {Ontological models and the interpretation of contextuality}.
\newblock {\em arXiv:}0709.4266, 2007.

\bibitem{Harrigan2007b}
N.~Harrigan, T.~Rudolph and S.~Aaronson.
\newblock {Representing probabilistic data via ontological models}.
\newblock {\em arXiv:}0709.1149, 2008.

\bibitem{How89} 
D.~Howard.
\newblock `Nicht sein kann was nicht sein darf,' or the Prehistory of EPR, 1909--1935: Einstein's Early Worries about the Quantum Mechanics of Composite Systems.
\newblock In {\em Sixty-Two Years of Uncertainty: Historical, Philosophical, and Physical Inquiries into the Foundations of Quantum Mechanics}, Proceedings of the 1989 Conference,``Ettore Majorana'' Centre for Scientific Culture, International School of History of Science, Erice, Italy, 5--14 August. Arthur Miller, ed. New York: Plenum, pp.~61--111.

\bibitem{Masanes2006General}
L.~Masanes, A.~Ac\'{i}n, and N.~Gisin.
\newblock {General properties of nonsignaling theories}.
\newblock {\em Physical Review A}, 73(1):012112, 2006.

\bibitem{Pironio2010Random}
S.~Pironio, A.~Ac\'{i}n, S.~Massar, A.~Boyer de~la Giroday, D.~N. Matsukevich,
  P.~Maunz, S.~Olmschenk, D.~Hayes, L.~Luo, T.~A. Manning, and C.~Monroe.
\newblock {Random numbers certified by Bell's theorem}.
\newblock {\em Nature}, 464:1021--1024, 2010.

\bibitem{Popescu1994b}
S.~Popescu and D.~Rohrlich.
\newblock {Quantum Nonlocality as an axiom}.
\newblock {\em Foundations of Physics}, 24(3):379--385, 1994.

\bibitem{Rudolph2006}
T.~Rudolph.
\newblock Ontological models for quantum mechanics and the {K}ochen-{S}pecker
  theorem.
\newblock {\em arxiv:}quant-ph/0608120, 2006.

\bibitem{Sch49}
P.~ A.~ Schilpp (ed.).
\newblock {\em {Albert Einstein: Philosopher-Scientist}}.
\newblock Library of the Living Philosophers, Evanston, 1949.

\bibitem{Shi89} 
A.~Shimony. 
\newblock In J.~T.~Cushing and E.~McMullin, (eds.). 
\newblock {\em Philosophical Consequences of Quantum Theory}, pp.~25--37. 
\newblock University of Notre Dame Press, Notre Dame, 1989. 

\bibitem{Spekkens2005}
R.~W. Spekkens.
\newblock {Contextuality for preparations, transformations, and unsharp
  measurements}.
\newblock {\em Physical Review A}, 71(5):052108, 2005.

\bibitem{Valentini2002}
A.~Valentini.
\newblock {Signal-locality in hidden-variables theories}.
\newblock {\em Physics Letters A}, 297:273--278, 2002.

\bibitem{QTMeas1983}
J.~A. Wheeler and W.~H. Zurek, editors.
\newblock {\em {Quantum Theory and Measurement}}.
\newblock Princeton University Press, 1983.

\end{thebibliography}

}
\end{document}